# Binary progenitor systems for Type Ic supernovae

Martín Solar ¹ ✉, Michał J. Michałowski¹ ✉, Jakub Nadolny ¹,
Lluís Galbany ²,³, Jens Hjorth⁴, Emmanouil Zapartas ⁵, Jesper Sollerman ⁶,
Leslie Hunt ⁷, Sylvio Klose⁸, Maciej Koprowski ⁹, Aleksandra Leśniewska¹,⁴,
Michał Małkowski ¹, Ana M. Nicuesa Guelbenzu⁸, Oleh Ryzhov¹,
Sandra Savaglio ¹⁰,¹¹,¹², Patricia Schady¹³, Steve Schulze ¹⁴,
Antonio de Ugarte Postigo¹⁵, Susanna D. Vergani¹⁶,¹⁷, Darach Watson ¹⁸,¹⁹ &
Radosław Wróblewski¹

Core-collapse supernovae are explosions of massive stars at the end of their evolution. They are responsible for metal production and for halting star formation, having a significant impact on galaxy evolution. The details of these processes depend on the nature of supernova progenitors, but it is unclear if Type Ic supernovae (without hydrogen or helium lines in their spectra) originate from core-collapses of very massive stars (>30 $M_\odot$) or from less massive stars in binary systems. Here we show that Type II (with hydrogen lines) and Ic supernovae are located in environments with similar molecular gas densities, therefore their progenitors have comparable lifetimes and initial masses. This supports a binary interaction for most Type Ic supernova progenitors, which explains the lack of hydrogen and helium lines. This finding can be implemented in sub-grid prescriptions in numerical cosmological simulations to improve the feedback and chemical mixing.

Massive stars (>8 $M_\odot$) have a significant impact on the interstellar medium (ISM) by regulating it through stellar winds, ionising radiation, and supernova (SN) explosions. SNe contribute to the origin of heavy elements, a still poorly understood aspect in cosmology¹. The details of SN feedback and metal production depend primarily on which stars explode as which type of SNe because the mechanism of explosion and element production yield is different for each of them. SN feedback models that take into consideration progenitor stars and mechanisms of explosions are essential to improve simulations.

The connection between SN types and their progenitors is of great importance for galaxy evolution and cosmology. This can be directly studied by identifying progenitor stars on pre-explosion images, with follow-up observations that confirm their disappearance. However, to date, only 23 (18 Type II, four Type IIb, and one Type Ib) core-collapse

¹Astronomical Observatory Institute, Faculty of Physics, Adam Mickiewicz University, Poznań, Poland. ²Institute of Space Sciences (ICE, CSIC), Campus UAB, Barcelona, Spain. ³Institut d'Estudis Espacials de Catalunya (IEEC), Barcelona, Spain. ⁴DARK, Niels Bohr Institute, University of Copenhagen, Copenhagen N, Denmark. ⁵Institute of Astrophysics, FORTH, Heraklion, Greece. ⁶The Oskar Klein Centre, Department of Astronomy, Stockholm University, Albanova University Center, Stockholm, Sweden. ⁷INAF—Osservatorio Astrofisico di Arcetri, Firenze, Italy. ⁸Thüringer Landessternwarte Tautenburg, Tautenburg, Germany. ⁹Institute of Astronomy, Faculty of Physics, Astronomy and Informatics, Nicolaus Copernicus University, Toruń, Poland. ¹⁰Department of physics, University of Calabria, Arcavacata di Rende (CS), Italy. ¹¹INAF—Osservatorio di Astrofisica e Scienza dello Spazio, Bologna, Italy. ¹²INFN—Laboratori Nazionali di Frascati, Frascati, Italy. ¹³Department of Physics, University of Bath, Bath, UK. ¹⁴The Oskar Klein Centre, Department of Physics, Stockholm University, Albanova University Center, Stockholm, Sweden. ¹⁵Artemis, Observatoire de la Cotê d'Azur, Université Cotê d'Azur, Nice, France. ¹⁶GEPI, Observatoire de Paris, PSL University, CNRS, Meudon, France. ¹⁷Institut d'Astrophysique de Paris, UMR 7095, CNRS-SU, Paris, France. ¹⁸The Cosmic Dawn Center (DAWN), Copenhagen, Denmark. ¹⁹Niels Bohr Institute, University of Copenhagen, Copenhagen N, Denmark. ✉e-mail: martin.solar@amu.edu.pl; michal.michalowski@amu.edu.pl

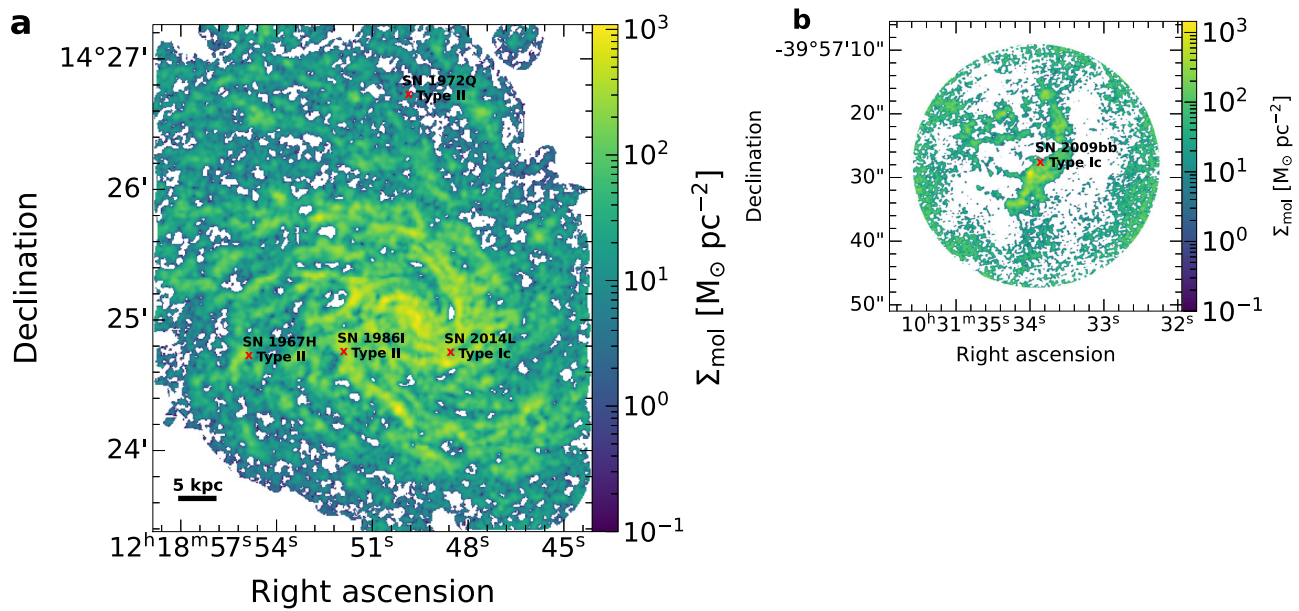

**Fig. 1 | Distribution of molecular gas in galaxy host and SN locations.** SN 1967H, SN 1972Q and SN 1986I hosted in NGC 4254 (**a**), and SN 2009bb hosted in NGC 3278 (**b**). Colour-coded $\Sigma_{mol}$ intensity is represented in logarithmic scale. Pixels without signal are masked and shown as white. Red dots represent SN locations. North is up and East is left.

SN (CCSN) progenitor stars have been confirmed to disappear in post-explosion images[2]. No Type Ic SN (without hydrogen or helium in the spectrum) progenitor has been confirmed in this way, with SN 2017ein as the only candidate, but with a wide range of derived progenitor masses[3–6]. Hence, most of our knowledge on the nature of Type Ic SN progenitors is based on photometric and spectroscopic observations after the explosion.

Type Ic SNe likely originates either from core-collapses of very massive stars or from less massive stars in binary systems. The mechanism responsible for the lack of hydrogen or helium lines is still a subject of debate. In the case of the very massive star model, the hydrogen and helium envelope is lost due to stellar winds from the star itself[7,8] (e.g., radiatively driven winds, episodic mass-loss, or rapid rotation). In the binary model, the companion radically affects the evolution of the progenitor due to the mass exchange[9–11]. Non-detections of Type Ic SN progenitors in pre-explosion images and relative rates of different types of SNe suggest that most of Type Ic SN progenitors are binaries with initial masses < 20 $M_\odot$[12,13].

Another way to address the relation between progenitor stars and resulting SNe is to investigate the molecular gas properties at the explosion location. Molecular gas at the locations of SNe of different types was recently investigated at a spatial resolution comparable to giant molecular clouds (GMCs)[14], using the Atacama Large Millimetre Array (ALMA) carbon monoxide 2-1 line transition [CO(2-1)] observations from the Physics at High Angular resolution in Nearby GalaxieS (PHANGS)[15,16] survey. Their sample consisted of a total of 59 SNe: 12 thermonuclear (Type Ia SNe), 32 Type II SNe, eight stripped-envelope SNe (SESNe, hereafter, Type Ib, Ic or Ib/c), and seven unclassified. They found that Type Ia and II SNe are associated with little or no molecular gas emission, while SESNe and unclassified SNe mostly show strong molecular gas emission. They concluded that there is a clear dependence of the type of SN and the molecular gas environment, however, their conclusions are drawn based on a low sample size for SESNe and, thus are not statistically significant.

In this work, our goal is to constrain lifetimes and initial masses of Type Ic SN progenitors. To this end, we compare the molecular gas densities at the positions of Type II and Ic SNe. By targeting a large sample of SNe we aim to uncover their nature. This statistical approach offers strong constraints on the overall progenitor characteristics of different SN populations but does not provide a strong constraint on individual SN progenitor properties. We report a statistically significant study to do so with spatial resolution comparable to the GMC sizes. This is an important factor because molecular hydrogen column surface densities and lifetimes of GMCs can only be measured accurately if the resolution at least matches the cloud sizes[17,18].

## Results

In order to investigate the environments of a significant number of SNe at high resolution, we initiated the ALMA CO SN (ACOS) survey, obtaining CO(2-1) observations of the locations of 16 Type Ic SNe. Together with the PHANGS survey this results in a sample of 63 SNe: 12 Type Ia, 30 Type II, and 21 Type Ic SNe. These CO(2-1) observations have a spatial resolution of ~100 pc, similar to sizes of GMCs. The spatial resolution and the large sample allow us to study the immediate environments in which the SNe exploded. Our main conclusion is from the comparison of Type II and Ic SNe, whereas Type Ia SNe are shown only to contrast the different progenitor natures. See 'Methods', subsections ALMA CO SN survey, PHANGS–ALMA data, and Supernova sample for detailed information about the SNe and their host galaxies. Supplementary Data 1 lists the information for the SNe used in this work.

As an example, Fig. 1 shows the molecular hydrogen gas surface density ($\Sigma_{mol}$) map of NGC 4254 (M 99, a typical PHANGS–ALMA galaxy) with its four CCSNe; three Type II (SN 1967H, SN 1972Q, and SN 1986I) and one Type Ic (SN 2014L) SNe, plus the location of SN 2009bb (Type Ic SN) hosted in NGC 3278 (observed by ACOS). $\Sigma_{mol}$ is computed from the CO(2-1) line intensity, see methods, subsection CO-to-$\Sigma_{mol}$ for the description used.

The $\Sigma_{mol}$ value for a given SN was calculated in two ways. First, we measured it at the exact pixel of the SN location and this is denoted as "SN location". However, the SN location might not be the exact site where the progenitor star formed. The true location of the formation of the SN progenitor star could be shifted due to astrometric displacement and/or peculiar motion of the progenitor system with respect to the parent GMC. To take into account these effects, together with the maximum size of GMCs, we also measured the maximum value within a radius of 200 pc centred at the SN position (see methods, subsection 200 pc regions). To assess an average level of $\Sigma_{mol}$ for our host galaxies, we also measured $\Sigma_{mol}$ for $10^3$ random pixels in each PHANGS galaxy and the respective maximum value within a radius of 200 pc centred at these random pixels. The ACOS sample was not included in the random pixel calculations because these observations cover the SN positions but not the whole galaxy, so it is not possible to obtain a representative measurement of $\Sigma_{mol}$ for these galaxies.

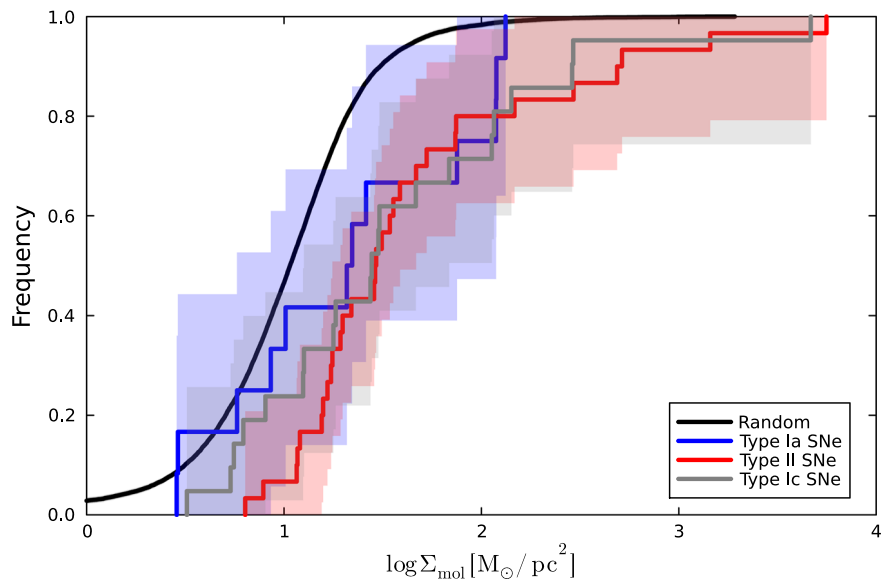

**Fig. 2 | $\Sigma_{mol}$ eCDFs for SN locations.** Random locations, Type Ia, Type II, and Type Ic SNe are represented by black, blue, red, and grey lines, respectively. Upper limits ($2\sigma$) are used in case of non-detections. The shaded areas represent confidence intervals at $1\sigma$.

Figure 2 shows the empirical cumulative distribution function (eCDF) of $\Sigma_{mol}$ in the environments of SNe with their respective confidence intervals calculated using the Dvoretzky–Kiefer–Wolfowitz inequality[19] at a significance level of 68%. If a given $\Sigma_{mol}$ measurement has a significance lower than $2\sigma$ (non-detection), where $\sigma$ is the noise at the pixel location, then the $2\sigma$ upper limit is shown on the eCDF. This happens for 66% of the random pixels (15077 out of 23000), 67% of Type Ia (eight out of 12), 37% Type II (11 out of 30) and 43% Type Ic SNe (nine out of 21). These upper limits are not used for any calculations, only for the visualisation of the eCDFs. Supplementary Data 2 and 3 lists the $\Sigma_{mol}$ environments of SNe and random locations in galaxy hosts, respectively. Random pixel locations extend towards lower values when compared to the SN positions, indicating that the SN locations have a strong connection with their GMC parents or larger molecular structures. This is an expected result for CCSNe, since they are associated with massive progenitor stars and therefore they are expected to explode close to their birthplaces in spiral arms where molecular gas densities are higher[20].

To quantify if our samples are drawn from different parent GMC populations, we performed a Kolmogorov−Smirnov (KS) test for each combination. KS statistics and $p$-values are shown in Table 1 (and for the 200 pc regions in Supplementary Table 1). Comparing the locations of CCSNe and random pixels, the low $p$-values indicate that their distributions are different, suggesting high densities of molecular gas environments for the sites where SNe was observed. On the other hand, high $p$-values (> 0.05) for SN combinations indicate that it is not possible to reject the null hypothesis that the $\Sigma_{mol}$ at the SN positions are from the same distribution.

In order to obtain confidence interval ranges of $\Sigma_{mol}$ for each SN type and random locations, we performed Monte Carlo simulations. Each measurement of an individual SN was perturbed according to its uncertainty, which was assumed to have a Normal distribution. We created $10^4$ of such perturbed sets, each time calculating its median value. The uncertainty was adopted to be at 16% and 84% of the simulated distribution. We show the results of these simulations in Fig. 3. The medians and $1\sigma$ confidence intervals for SN locations and 200 pc regions are summarised in Supplementary Table 2.

Median molecular gas densities increase from the values measured for the random pixels ($4.47^{+0.05}_{-0.04}$ $M_\odot$ pc$^{-2}$), through Type Ia SNe ($6.93^{+3.70}_{-2.36}$ $M_\odot$ pc$^{-2}$), to Type II and Ic SNe ($20.15^{+3.38}_{-2.46}$ and $20.62^{+4.28}_{-4.88}$ $M_\odot$ pc$^{-2}$, respectively). The results are shown in Fig. 3. The random locations and Type Ia SNe have much lower median molecular gas densities than CCSNe. At the positions of Type II and Ic SNe we obtained similar molecular gas densities within the $1\sigma$ confidence levels.

Under the assumption of single very massive star progenitors for Type Ic SNe (for masses of 30–100 $M_\odot$ with lifetimes of 7–3 Myr, respectively[21]), it is expected that the respective parent GMC would not have been dispersed before the SN explosion due to a short progenitor lifetime, and the progenitors would not have enough time to shift away from their birthplaces significantly. Therefore, if progenitors of Type Ic SNe were very massive stars, then the distribution of $\Sigma_{mol}$ at their positions should be shifted toward higher values compared to that of Type II SNe. This is because the lower masses of progenitors of Type II SNe imply longer lifetimes, and therefore more time for the parent clouds to disperse and for the progenitors to move away. This scenario is not supported by our results. In the alternative scenario, the progenitors of Type II SNe evolve as single stars (or in wide binaries in which their hydrogen layers are not affected) and those of Type Ic SNe are similarly massive stars that evolve in binary systems with a companion being responsible for removing the external layers of hydrogen and helium[21]. Then the progenitor masses, and therefore lifetimes, of both types are similar. Thereby, their distributions of $\Sigma_{mol}$ should be comparable, which is consistent with our data. See methods, subsection Timing the SN progenitor lifetime with molecular gas information for the justification of using the molecular gas densities to constrain the stellar population age.

**Table 1 | KS statistics (and $p$-value in parenthesis) for different SN groups (SN locations)**

|        | Random          | Ia               | II               | Ic               |
|--------|-----------------|------------------|------------------|------------------|
| Random | 0.0 (1.0)       | 3.1e-01 (1.6e-01)| 4.5e-01 (4.7e-06)| 4.1e-01 (1.0e-03)|
| Ia     |                 | 0.0 (1.0)        | 4.0e-01 (1.0e-01)| 2.4e-01 (7.1e-01)|
| II     |                 |                  | 0.0 (1.0)        | 2.5e-01 (3.7e-01)|
| Ic     |                 |                  |                  | 0.0 (1.0)        |

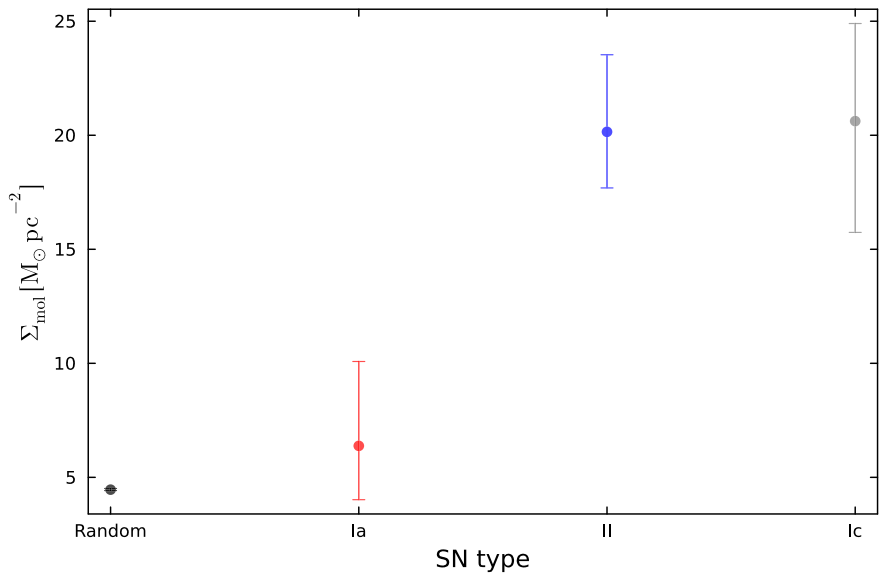

**Fig. 3 | Median values and 1σ confidence intervals (using $10^4$ Monte Carlo simulations) of $\Sigma_{mol}$ for SN locations.** Random locations, Type Ia, Type II, and Type Ic SNe are represented by black, blue, red, and grey error bars, respectively.

We note that at lower gas densities the distribution of Type Ic SNe is slightly lower than that of Type II SNe (see Fig. 2). This may indicate that the lifetimes of some of the Type Ic progenitors were increased by the binary interaction[21]. However, the statistical significance of this difference is too low to draw any definitive conclusions.

In order to assess the maximum difference between the lifetimes of progenitors of Type II and Ic SNe, we assumed that the molecular gas density at the SN progenitor positions, $\Sigma_{mol,SN}$, decreases exponentially with time as

$$\Sigma_{mol,SN} = \Sigma_0 \, e^{-t_{SN}/\tau_{GMC}}, \qquad (1)$$

where $\Sigma_0$ is a normalisation constant, $t_{SN}$ is the lifetime of the SN progenitor and $\tau_{GMC}$ is the lifetime of the GMC. In order to constrain the lifetimes of Type Ic SN progenitors, we made use of our measured ratio of $\Sigma_{mol,SN}$ for Type Ic and II SNe ($\Sigma_{mol,Ic}/\Sigma_{mol,II} = 1.02^{+0.27}_{-0.27}$) and a measured value of the GMC lifetime of $\tau_{GMC} = 16^{+6}_{-5}$ Myr[22].

This characteristic cloud evolution lifetime is in agreement with theoretical works and simulations[18,22–27] and while the exact value of the assumed average GMC lifetime influences these calculations, it does not change the interpretation when lifetimes of two SN types are compared. Assuming that progenitors of Type II and Ic SNe are born in GMCs with similar average initial conditions, i.e. that average $\Sigma_0$ is the same for both, from eq. (1) it is possible to calculate the difference between the SN progenitor lifetimes as

$$t_{II} - t_{Ic} = \tau_{GMC} \ln(\Sigma_{mol,Ic}/\Sigma_{mol,II}) = 0.37^{+4.27}_{-4.26} \text{ Myr.} \qquad (2)$$

A zero-age main sequence (ZAMS) mass $M_{init,SN}$ for Type II SN progenitors of $M_{init,II} = 10.66^{+0.20}_{-0.20}$ M$_\odot$ (obtained by averaging nine SNe with pre-explosion detection[28] and confirmed by the disappearance in post-explosion images[2]) yields a lifetime of $t_{II} = 25.22^{+0.80}_{-0.80}$ Myr. Using this $t_{II}$ in eq. (2) results in a lifetime for Type Ic SN progenitors of $t_{Ic} = 24.9^{+4.3}_{-4.3}$ Myr and a ZAMS mass of $M_{init,Ic} = 10.90^{+1.20}_{-1.20}$ M$_\odot$. On the other hand, if we assume a typical mass for red supergiant progenitors for type II SNe[29] of $M_{init,II} = 15^{+1}_{-1}$ M$_\odot$, then we obtain $M_{init,Ic} = 15.3^{+3.2}_{-3.2}$ M$_\odot$. This also means that if Type II SN progenitors include rare examples of very massive stars, so can Type Ic SN progenitors[30]. To account for signal unrelated to SNe, as the first step we also subtracted the random values of $\Sigma_{mol}$ from those of SNe, which resulted in $t_{II} - t_{Ic} = 0.47^{+5.47}_{-5.45}$ Myr, $t_{Ic} = 24.8^{+5.5}_{-5.5}$ Myr, and $M_{init,Ic} = 10.9^{+1.5}_{-1.5}$ M$_\odot$, indistinguishable from the original results.

Another effect to take into account is that SN progenitors may be runaway stars, which are moving away from their parent clusters with significant velocities. Maximum velocities for OB runaway stars are ~ 30 km s$^{-1}$ or ~ 30 pc/Myr[31], so they would need ~ 3 Myr to cross a GMC. Replacing $\tau_{GMC}$ with the effective crossing timescale $\tau_{cro}$, Eq. (1) will then be smaller, including both the movement of the progenitors and the cloud dispersal. Using $\tau_{cro} = 3$ Myr, Eq. (2) yields an even smaller lifetime difference between Type II and Ic SNe of ~ 0.06 Myr, what makes our conclusions even stronger.

If most of Type Ic SN progenitors were very massive stars with masses around 30 M$_\odot$, then for their lifetime of 7 Myr, Eq. (1) results in molecular gas densities a factor of 4 higher than those at the positions of Type II SNe, which is not in agreement with our results.

## Discussion

Our findings indicate that the binary interaction model (mass transfer due to a companion) is the main mechanism extracting outer layers for most Type Ic SN progenitors. However, we remark that we do not reject the possibility that strong stellar winds of a high-mass star can blow away the layers of hydrogen and helium leading to an explosion as a Type Ic SN, individual Type Ic SNe can be due to very massive star progenitors[32]. Indeed our calculations show that the 1σ range of the accepted fraction of low-mass stars in the Type Ic SNe population is 66–100% (see methods, subsection Statistical significance of the sample). Another possible binary progenitor scenario includes mergers or accretion of significant mass from a companion so that an initially low-mass star becomes massive and luminous enough to launch strong winds shedding the outer envelope before explosion.

Our results are consistent with low measurements of Type Ic SN progenitor masses from light curve modelling[33], the comparison of the SN rates[12], the modelling of emission lines at the positions of SNe Type Ib/c[29], and the direct observational evidence of a binary system for Type Ic SN 2022jli[34]. Moreover, in our Galaxy, ~ 70% of massive O-type stars are formed in close binary systems and are expected to experience mass transfer during their lifetime[35]. All these works support our conclusion about binary systems as progenitors of Type Ic SNe.

On the other hand, our low progenitor masses for Type Ic SNe are inconsistent with some other measurements. The $^{56}$Ni yields of Type Ic SNe are five times higher than those of Type II SNe[36]. However, this may not necessarily imply higher progenitor masses, but different explosion mechanisms or energy sources[37]. If Type Ic SN explosions are asymmetric or not predominantly powered by the radioactive decay of $^{56}$Ni, then the nickel mass will be overestimated for them.

Moreover, using PHANGS data, ref. 14 reported that SESNe (Type IIb, Ib, Ib/c, and Ic SNe) are associated with stronger CO(2-1) emission than those for Type Ia and II SNe. Their sample included 59 SNe in 28 galaxies, with the Type Ic sample size being the main difference compared to this work (they included only five of them, see methods, subsection Statistical significance of the sample for a statistical comparison).

SESNe (including Type Ic SNe) are associated with higher H$\alpha$ intensities than Type II SNe, suggesting younger star-forming regions and higher progenitor masses[38–44]. However, these results are complicated by the fact that the H$\alpha$ emission disappears on a timescale of several Myr only for isolated star-forming regions, whereas for larger complexes (several hundreds of pc, as probed by these observations), this timescale may be as long as 20 Myr[45]. Low ages for Type Ic SN locations were also obtained from population fitting of stars in the SN vicinity[46,47], but the ages may be underestimated due to blending of stars[47]. Moreover, stronger associations of Type Ic SNe with H$\alpha$ and UV-bright stars may also be explained within a scenario in which their lower-mass progenitors prefer regions of high stellar density or more top-heavy initial mass function (IMF), which increase the binarity fraction, but also the H$\alpha$ and UV emission. High-resolution multi-wavelength observations are needed to alleviate this tension.

Broad-lined Type Ic (Type Ic-BL) SNe and gamma-ray bursts, both known to be connected with very massive progenitor stars[48], show twice the amount of synthesised $^{56}$Ni than regular Type Ic SNe[49]. Our sample contains only four Type Ic-BL SNe and when they are excluded from the Type Ic SN sample, the results remain the same (at the 1$\sigma$ level). Similarly, when Type IIn and IIn/LBV (2 in total) are removed from the Type II SN sample, the results do not change.

The lack of statistically significant difference between Type Ia and II SN positions could be due to a low number of statistics for the former. Subtracting the median molecular density at the random positions from that of Type Ia SNe yields a non-detection: $\Sigma_{mol} = 1.93^{+3.70}_{-2.36}$ M$_\odot$ pc$^{-2}$. Using a corresponding 2$\sigma$ upper limit in Eq. (2) gives a lower limit on the lifetime difference between the Type Ia and II SN progenitors of $t_{Ia} - t_{II} > 12$ Myr, consistent with the expected higher lifetimes of Type Ia SN progenitors.

Implementation of our findings into cosmological simulations will have important impact on our understanding of the feedback processes. The progenitor type can be implemented in sub-grid prescriptions in numerical cosmological simulations[50–52] with regards to SN feedback and chemical yields into the ISM. Moreover, whether the progenitors of Type Ic SNe are binary systems or very massive stars changes their contribution to the formation of heavy elements, one of the key aspects of stellar and galaxy evolution[1]. This includes carbon, one of the most fundamental elements in the Universe and building block of life. At solar metallicity, within the Local Universe (including the Milky Way), stellar winds and SN explosions from binary–stripped stars are found to produce twice more $^{12}$C than similar single stars[53]. The fourth most abundant element in the Solar system is carbon (after hydrogen, helium, and oxygen) and a significant contribution could be produced from Type Ic SNe. This method can also be applied to larger samples divided into different properties (e.g., explosion characteristics, host galaxy types, environmental metallicities, etc) and more rare events to learn about their nature.

## Methods
### ALMA CO SN survey
ACOS (ALMA ID 2021.1.00099.S, P.I. M.J.M.) consists in observations of the $J = 2 \to 1$ transition of the $^{12}$CO line in the environment of 16 Type Ic SNe with an angular spatial resolution range of 0.4–1.1″ so that the physical resolution is around 50–100 pc. The selection criteria were the observability with ALMA, i.e. declination < 20°, and distances < 55 Mpc (redshift $z$ < 0.013) to allow the detection of individual GMCs in a reasonable observing time. We excluded seven hosts which are edge-on, for which projection effects would make it difficult to measure the gas surface density at the SN position. The distances, masses and luminosities of these galaxies are comparable to PHANGS–ALMA galaxies.

### PHANGS–ALMA data
PHANGS survey provides CO(2-1) line observations using ALMA for 74 galaxies in the Local Universe (<20 Mpc), which mostly are face-on ($i < 75^\circ$). The typical resolution was ~ 2″, corresponding to ~ 100 pc comparable to the sizes of GMCs. We used calibrated data from ref. 15 for the PHANGS–ALMA galaxy sample[16]. Further information about these procedures can be found in refs. 15,16.

### Supernova sample
Our SN sample was compiled from the Open Catalogue for Supernova (https://github.com/astrocatalogs/supernovae/) in April 2022[54]. The SN compilation, as designated in the catalogue, consists of Type Ia (Ia, Ia-02cx, and IaPec), Type II (II, IIP, IIn, and IIn/LBV), and Type Ic (Ic and Ic-BL). We did not consider Type I, Ib/c, IIb or SNe without classification because they cannot be categorised as any of our well-defined thermonuclear and CCSNe. Only two Type Ib SNe exploded in PHANGS galaxies, so they were also excluded from the analysis. We obtained a total of 63 SNe hosted in 39 galaxies (either single or multiple SNe located in one galaxy): 16 Type Ic SNe within ACOS and 47 SNe in 23 PHANGS galaxies. Out of these 47 SNe, 12 were thermonuclear Type Ia, 30 Type II and five Type Ic. We note that the sample may be biased by observational limitations, so it can miss dust-embedded SNe and those in the outskirts of the PHANGS galaxies not covered by the ALMA data. We also do not constrain SNe that do emerge from very massive stars that have not been able to escape from their parent GMCs but are so extinguished that they do not make it into an optical sample. This can only be addressed by an infra-red-selected SN sample. We also do not constrain the properties of stars which collapse directly to black holes (BHs)[55], because they do not feature in our SN samples. However, their exclusion does not affect significantly our results (see methods, subsection Statistical significance of the sample).

### CO-to-$\Sigma_{mol}$ conversion
The CO(2-1) velocity-integrated intensities from moment-0 maps were converted to $\Sigma_{mol}$ using the following equation (eq. 10 from ref. 16):

$$\Sigma_{mol} = \alpha_{CO}^{1-0} R_{21}^{-1} I_{CO(2-1)} \cos i, \qquad (3)$$

where $\alpha_{CO}^{1-0}$ is the CO(1-0) conversion factor, $R_{21}$ is the CO(2-1)-to-CO(1-0) line ratio, $i$ is the inclination angle of the galaxy, and $I_{CO(2-1)}$ is the line-integrated CO(2-1) intensity. We adopt a Galactic CO-to-H$_2$ conversion factor of $\alpha_{CO}^{1-0} = 5$ M$_\odot$ pc$^{-2}$ (Kkms$^{-1}$)$^{-1}$, the same as in ref. 56, and a line ratio of $R_{21} = 0.5$, from ref. 57. Inclination angles were taken from ref. 58 for the PHANGS sample, and from the Hyperleda (http://leda.univ-lyon1.fr/) galaxy database[59] for ACOS galaxy hosts.

### 200 pc regions
An SN explosion could be located away from the centre of a cloud. In order to have a better understanding of the parent GMCs, the maximum value $\Sigma_{mol}$ in a circumference within a radius of 200 pc centred on the SN position was also calculated and denoted as "200 pc region".

In the Milky Way, this radius is comparable to the maximum GMC size[60]. Supplementary Fig. 1 shows the $\Sigma_{mol}$ eCDFs in such 200 pc regions, with a clear shift to higher densities compared with SN locations, as expected. The two-sample KS test from Supplementary Table 1 shows high probabilities that each of the location pairs is drawn from the same distribution. The median and $1\sigma$ values obtained via Monte Carlo simulations are shown in Supplementary Fig. 2. The fact that Type II SNe reach molecular gas densities higher than Type Ic SNe strengthens our conclusions.

### Cosmological model

We use the nine-year Wilkinson Microwave Anisotropy Probe cosmological model[61] with parameters $H_0 = 69.32$ km s$^{-1}$ Mpc$^{-1}$, $\Omega_\Lambda = 0.7134$, and $\Omega_m = 0.2865$. Redshift values were used only to compute the 200 pc region for each SN and have no influence on the physical interpretation of the results.

### Timing the SN progenitor lifetime with molecular gas observations

The use of molecular gas density as a tracer of stellar age is justified by the strong correlation between the age distribution and the cluster-GMC distance[62–64]. Moreover, there is a close association between the birth environment (i.e. GMC separation) and age of the cluster, measured by the equivalent width (EW) of H$\alpha$ [EW(H$\alpha$)][65]. Finally, the analysis of stellar associations and the CO(2-1) emission revealed that the percentage of overlap between the regions of stellar associations and GMCs is ~60%[66].

In order to test if there is a correlation between molecular gas densities and stellar ages in the PHANGS sample, a pixel-to-pixel comparison was computed for $\Sigma_{mol}$ and EW(H$\alpha$), a proxy for age. The H$\alpha$ maps were obtained from the Multi Unit Spectroscopic Explorer (MUSE)[67]. The continuum maps were collected from the Wide Field Imager (La Silla's 2.2m MPG/ESO telescope)[68] and also available in the PHANGS−MUSE dataset.

Supplementary Fig. 3 shows the relation of $\Sigma_{mol}$ and EW(H$\alpha$) for every pixel of our galaxy sample, i.e. 11 PHANGS galaxies (NGC 628, NGC 1087, NGC 1365, NGC 1385, NGC 1433, NGC 1566, NGC 1672, NGC 3627, NGC 4254, NGC 4303, and NGC 4321), with both ALMA and MUSE data, and hosting at least one SN from our sample. There is a clear correlation with pixels with lower EW(H$\alpha$) (older) having lower molecular gas density, The scatter is significant, but we take the scatter of this magnitude into account in our significance test below (and this justifies the need of a sample of the order of a few tens of SNe).

### Statistical significance of the sample

To assess the statistical significance of our results with respect to the sample size, we generated $10^4$ sets of synthetic parent GMC densities for 30 Type II SNe (as in our data) and a variable number of very massive stars in order to test if we can distinguish them. We have done it in three ways, first starting from the measured gas densities of Type II SNe (method 1), second starting from the measured gas density distribution in PHANGS galaxies (method 2), and last from lifetimes of binary systems from a numerical model (method 3).

For the former case, in order to have a realistic distribution of GMC densities we need to remove the outliers of $\Sigma_{mol,II}$ data because their high values do not correspond to densities of single GMCs (as we intend to probe), but the accumulation of GMCs along the line-of-sight towards to galaxy centres, where indeed, all identified outliers are located. We obtained the first, second, and third quartiles of $\Sigma_{mol,II}$ ($Q_1$, $Q_2$, and $Q_3$, respectively) and considered outliers as values lower than $Q_1 - 1.5 * IQR$ or higher than $Q_3 + 1.5 * IQR$, where $IQR = Q_3 - Q_1$ is the interquartile range. After removing outliers (292, 486, 515, 1442, and 5599 M$_\odot$pc$^{-2}$, higher than $Q_3 + 1.5 * IQR = 157$ M$_\odot$ pc$^{-2}$), we found an analytical function which best reproduces the distribution of Type II SN $\Sigma_{mol}$ locations by fitting around ~80 different distributions[69]. The best function was a generalised normal continuous random distribution $f(x, \beta) = \frac{\beta}{2\Gamma(1/\beta)} e^{-|x|^\beta}$, where $x$ is a real number, $\beta > 0$ is the shape parameter, and $\Gamma$ is a gamma function. The fitted parameters were $\beta = 0.51$, centred at 11.7 with a scale of 3.52. From this distribution, we constructed two different synthetic distributions corresponding to Type II SNe and very massive stars to assess our ability to distinguish them. For Type II SNe we randomly drew from the function fitted above. For the very massive stars, we made use of Eq. (2) to derive their median $\Sigma_{mol,massive} = \Sigma_{mol,II} e^{(t_{II} - t_{massive})/\tau_{GMC}} = 4 \Sigma_{mol,II}$ and drew from a similar function scaled by this factor. In this calculation we assumed the initial mass for Type II SN progenitor of $M_{init,II} = 11$ M$_\odot$, corresponding to a lifetime of $t_{II} = 25$ Myr and an initial mass of $M_{init,massive} = 30$ M$_\odot$, corresponding to a lifetime of $t_{massive} = 3$ Myr. Finally, we assumed $\tau_{GMC} = 16$ Myr.

In the second method, for each SN we drew a random progenitor age from a normal distribution of $25 \pm 5$ Myr and $3 \pm 1$ Myr for Type II SNe and very massive stars, respectively, and the lifetime of the GMC of $\tau_{GMC} = 16 \pm 5$ Myr. We also drew an initial GMC gas density from a distribution with a mean value 0.5 dex higher than the observed distribution[70] (because the initial densities were higher for all of the observed clouds) and the same width, so that $\log(\Sigma_0 / M_\odot pc^{-2}) = 2.0 \pm 0.5$. The value of this parameter has no influence on the results, as this is only a normalisation and was chosen so that the median of the simulated distribution for Type II SNe is consistent with the observed value. Then we evolved the clouds as an exponential decay to calculate the surface densities at the time of the SN explosions (Eq. (1)).

For the third test, in order to take into account the effect of binarity in a simplified way, we drew samples of Type II SN progenitors and massive stars from the ranges of 6 M$_\odot$ − $M_{thresh}$ and $M_{thresh}$ − 100 M$_\odot$, respectively, for $M_{thresh} = 15, 20, 25,$ and 30 M$_\odot$, weighting with the Kroupa IMF. We randomly assigned an age according to the age probability distribution of SN progenitors for a given initial mass according to the models of ref. 21. Prior mass exchange of the progenitor with its companion leads in general to a longer lifetime. The two mass ranges represent tentative progenitors of Type II SNe and stripped-envelope SNe, although various binary scenarios violate this threshold. In a way, this test takes into account the change of lifetimes due to binarity, without accounting for a possible change in the SN type due to it.

For each method and for each simulated pair of sets (Type II SNe and very massive stars), we performed the KS test in order to check if we could reject the incorrect-by-design null hypothesis that they are drawn from the same distribution. Supplementary Fig. 4 shows the percentages of $p$-values below 0.05 (to reject the null hypothesis) and 0.37 (measured value from Table 1) for $10^4$ Monte Carlo simulations from a KS two-sample test between the distributions of the 30 random values of Type II SNe and the massive stars constructed above, as a function of sample size for such massive stars. With the sample size of 21, as in our sample of Type Ic SNe, in these simulations, in ~96% of the cases we obtained the $p$-value lower than 0.05 (and in 99.9% of cases lower than the measured value of 0.37). This means that we have statistical significance to correctly reject the null hypothesis and if Type Ic SNe were very massive stars, then we would obtain a lower $p$-value than we measured for virtually all the cases, so our data have enough statistical significance to rule out the very massive star hypothesis.

We also tested how the data can constrain a mixed Type Ic SN population, by analysing the fraction of the simulations with higher $p$-values than measured when we replaced some of the massive stars by lower mass progenitors in the same range as we assumed for Type II SNe. The $1\sigma$ range (68% of the simulated samples having a $p$-value higher than measured) of the accepted fraction of low-mass stars in the Type Ic SN population is 66–100%. Hence, only a third of the Type Ic SNe could be very massive stars, so that we could still measure the high $p$-value.

Moreover, in method 2, instead of drawing ages from normal distributions, we also drew masses according to the Kroupa IMFs[71] and calculated their lifetimes according to the relationship of ref. 21. In this case, for all the values of $M_{thresh}$ listed above, the number of the simulated samples having $p$-values lower than the measured value decreased from 99.9% to 97–98%. Finally, none of these calculations was significantly affected by the exclusion of the mass ranges for which no SNe are expected, due to a direct collapse into BHs, i.e. within the ranges 22–25 and 27–60 $M_\odot$[55]. If this is taken into account the significance increases by 1–2% due to making the difference between the Type II SNe and massive stars more pronounced.

**Lifetime–initial mass relation**

We converted the ZAMS masses to lifetimes using the lifetime–initial mass relation for single stars from ref. 21 (see their Fig. 1).

## Data availability

ACOS imaging is available from https://almascience.eso.org/aq/ under the proposal ID 2021.1.00099.S (P.I. M.J.M.). PHANGS–ALMA and PHANGS–MUSE imaging are available from https://www.canfar.net/storage/vault/list/phangs/RELEASES/ SNe information (name, astrometric positions, type, and redshift) were collected from https://github.com/astrocatalogs/supernovae/ to the date of April 2022. This paper makes use of the following ALMA data: ADS/JAO.ALMA#2012.1.00650.S, ADS/JAO.ALMA#2013.1.01161.S, ADS/JAO.ALMA#2015.1.00121.S, ADS/JAO.ALMA#2015.1.00925.S, ADS/JAO.ALMA#2015.1.00956.S, ADS/JAO.ALMA#2016.1.00386.S, ADS/JAO.ALMA#2017.1.00392.S, ADS/JAO.ALMA#2017.1.00886.L, ADS/JAO.ALMA#2018.1.01651.S, ADS/JAO.ALMA#2021.1.00099.S Based on data products created from observations collected at the European Organisation for Astronomical Research in the Southern Hemisphere under ESO programme(s) 1100.B-0651, 095.C-0473, and 094.C-0623 (PHANGS–MUSE; PI Schinnerer), as well as 094.B-0321 (MAGNUM; PI Marconi), 099.B-0242, 0100.B-0116, 098.B-0551 (MAD; PI Carollo) and 097.B-0640 (TIMER; PI Gadotti). Source data are provided with this paper. The Supplementary Data 1, 2, and 3 generated in this study have been deposited in the Zenodo database under accession code https://doi.org/10.5281/zenodo.13152457. Source data are provided with this paper.

## Acknowledgements

M.S., M.J.M., J.N., and A.L. acknowledge the support of the National Science Centre, Poland through the SONATA BIS grant 2018/30/E/ST9/00208. M.J.M. acknowledges the support of the Polish National Agency for Academic Exchange Bekker grant BPN/BEK/2022/1/00110. L.G. acknowledges financial support from the Spanish Ministerio de Ciencia e Innovación (MCIN), the Agencia Estatal de Investigación (AEI) 10.13039/501100011033, and the European Social Fund (ESF) "Investing in your future" under the 2019 Ramón y Cajal program RYC2019-027683-I and the PID2020-115253GA-I00 HOSTFLOWS project, from Centro Superior de Investigaciones Científicas (CSIC) under the PIE project 20215AT016, and the programme Unidad de Excelencia María de Maeztu CEX2020-001058-M. This work was supported by research grants (VIL16599, VIL54489) from VILLUM FONDEN. EZ acknowledges support from the Hellenic Foundation for Research and Innovation under the "3rd Call for H.F.R.I. Research Projects to Support Post-Doctoral Researchers" (Project No: 7933). O.R. acknowledge the support of the National Science Centre, Poland through the grant 2022/01/4/ST9/00037. This research was funded in whole or in part by the National Science Centre, Poland (grant numbers: 2020/39/D/ST9/03078 and 2021/41/N/ST9/02662). We acknowledge David Alex Kann who passed away before the submission of this manuscript and contributed to the writing of the observing proposal and interpretation of the data. The Cosmic Dawn Center (DAWN) is funded by the Danish National Research Foundation under grant DNRF140. D.W. is co-funded by the European Union (ERC, HEAVYMETAL, 101071865). Views and opinions expressed are, however, those of the authors only and do not necessarily reflect those of the European Union or the European Research Council. Neither the European Union nor the granting authority can be held responsible for them. ALMA is a partnership of ESO (representing its member states), NSF (USA) and NINS (Japan), together with NRC (Canada), MOST and ASIAA (Taiwan), and KASI (Republic of Korea), in cooperation with the Republic of Chile. The Joint ALMA Observatory is operated by ESO, AUI/NRAO and NAOJ. The National Radio Astronomy Observatory is a facility of the National Science Foundation operated under cooperative agreement by Associated Universities, Inc. The Joint ALMA Observatory is operated by ESO, AUI/NRAO and NAOJ. Based on observations taken as part of the PHANGS–MUSE large programme[67]. This research has made use of the services of the ESO Science Archive Facility. This research has made use of the services of the ESO Science Archive Facility. Science data products from the ESO archive may be distributed by third parties, and disseminated via other services, according to the terms of the Creative Commons Attribution 4.0 International license (https://creativecommons.org/licences/by/4.0/). Credit to the ESO origin of the data must be acknowledged, and the file headers preserved. This work is based [in part] on observations made with the Spitzer Space Telescope, which was operated by the Jet Propulsion Laboratory, California Institute of Technology under a contract with NASA. We acknowledge the usage of the HyperLeda database (http://leda.univ-lyon1.fr).

## Author contributions

M.S. performed most of the analysis and led writing of the manuscript. M.J.M. conceived the idea, led the ALMA proposal 2021.1.00099.S on which this work is based, and performed two of the significance tests. M.S., M.J.M., and J.N. coordinated the project. L.G., J.H., E.Z., and J.S. provided significant contributions to the interpretation of the data. L.G., J.H., L.H., S.K., M.K., A.L., M.M., A.M.N.G., Sandra S., P.S., Steve S., J.S., A.dUP., S.D.V., and D.W. contributed to the writing of the observing proposal. R.W. supported the data analysis and improved the text. M.M. and O.R. compiled the SN list. All the authors contributed to writing the manuscript.

## Competing interests

The authors declare no competing interests.

## Additional information

**Supplementary information** The online version contains supplementary material available at
https://doi.org/10.1038/s41467-024-51863-z.

**Correspondence** and requests for materials should be addressed to Martín Solar or Michał J. Michałowski.

**Peer review information** *Nature Communications* thanks the anonymous reviewers for their contribution to the peer review of this work. A peer review file is available.

**Reprints and permissions information** is available at
http://www.nature.com/reprints

**Publisher's note** Springer Nature remains neutral with regard to jurisdictional claims in published maps and institutional affiliations.